\documentclass[aps,pre,twocolumn,superscriptaddress,longbibliography,floatfix]{revtex4-2}
\usepackage{graphicx,bm,amssymb,amsmath,dcolumn,hyperref}
\usepackage{multirow}
\usepackage{enumerate}
\usepackage{enumitem}
\usepackage{color}
\usepackage{epstopdf}
\usepackage{bbm}
\usepackage{hyperref}
\usepackage{threeparttable}
\usepackage{amsthm}
\usepackage{bm}        
\usepackage{mathtools}
\usepackage{color}
\usepackage{tikz}
\usepackage{float}
\usepackage{empheq}
\usepackage[ruled,linesnumbered]{algorithm2e}
\definecolor{blue}{rgb}{0,0,1}
\definecolor{red}{rgb}{1,0,0}
\definecolor{green}{rgb}{0,1,0}

\begin{document}

\title{Universality of the complete-graph Potts model with $0< q \leq 2$}

\author{Zirui Peng}
\affiliation{Department of Modern Physics, University of Science and Technology of China, Hefei, Anhui 230026, China}

\author{Sheng Fang}
\affiliation{Hefei National Research Center for Physical Sciences at the Microscales, University of Science and Technology of China, Hefei, Anhui 230026, China}
\affiliation{School of Systems Science and Institute of Nonequilibrium Systems, Beijing Normal University, Beijing 100875, China}

\author{Hao Hu}
\email{Contact author: huhao@ahu.edu.cn}
\affiliation{School of Physics and Optoelectronic Engineering, Anhui University, Hefei, Anhui 230601, China}

\author{Youjin Deng}
\email{Contact author: yjdeng@ustc.edu.cn}
\affiliation{Department of Modern Physics, University of Science and Technology of China, Hefei, Anhui 230026, China}
\affiliation{Hefei National Laboratory, University of Science and Technology of China, Hefei, Anhui 230088, China}

\begin{abstract}
Universality is a fundamental concept in modern physics. 
For the $q$-state Potts model, the critical exponents are merely determined by the order-parameter symmetry $S_q$, 
spatial dimensionality and interaction range, independent of microscopic details. 
In a simplest and mean-field treatment—i.e., the Potts model on complete graph (CG), 
the phase transition is further established to be of percolation universality for the range of $0 < q <2$. 
By simulating the CG Potts model in the random-cluster representation, 
we numerically demonstrate such a hyper-universality that the critical exponents are the same for $0< q <2$ and, 
moreover, the Ising system ($q = 2$) exhibits a variety of critical geometric properties in percolation universality. 
On the other hand, many other universal properties in the finite-size scaling (FSS) theory, 
including Binder-like ratios and distribution function of the order parameter, are observed to be $q$-dependent. 
Our finding provides valuable insights for the study of critical phenomena in finite spatial dimensions, 
particularly when the FSS theory is utilized.
\end{abstract}
\maketitle

\section{\label{sec1}Introduction }

In phase transitions, 
such as the transition from a liquid to a gas state or from a ferromagnetic to a paramagnetic state, 
critical behavior describes properties of a system near its critical point.
Typical critical behavior is that physical observables follow power-law functions, whose exponents characterize the scaling of the observables.
Universality is a fundamental concept in understanding critical behavior. 
It refers to the idea that critical behavior of physical systems often depends only on a few essential characteristics, 
such as the spatial dimension, order-parameter symmetry and interaction range, regardless of microscopic details.
Researchers usually classify phase transitions into different universality classes~\cite{wilson1983renormalization,pelissetto2002critical}, 
each characterized by a set of critical exponents and other universal properties. 
For example, many systems with short-range interactions and a scalar order parameter undergo a transition belonging to the Ising universality class~\cite{pelissetto2002critical}, 
such as the liquid–vapor transition in simple fluids~\cite{parola1995liquid}, the phase separation in binary fluid mixtures~\cite{cloizeaux1991polymers}, 
the ferromagnetic-paramagnetic phase transition in magnetic materials, and the micellization transition in micellar systems~\cite{hall1972exact}.
Recent research work demonstrated the prevalence of Ising universality also in soft matter and active matter, e.g., soft-matter fluids in bulk~\cite{vink2009critical}, active Lennard-Jones fluids, and motility-induced phase separation in active systems~\cite{paoluzzi2020statistical}.
Understanding universality enables us to make predictions about critical behavior of systems based on general principles, without investigating details of each system. 

There exist various levels of universality for different quantities
(see e.g., Ref.~\cite{mertens2017universal} and references therein).
Critical exponents usually exhibit the strongest universality, in the sense that usually they
are not affected by the specific system (on-lattice or off-lattice, percolation in site type or bond type, etc.) 
or boundary conditions, and only depend on the spatial dimension, the interaction range and the order-parameter symmetry. Then, there are quantities whose scaling functions also do not depend on the specific system or boundary conditions, but need metric factors for near-critical data collapsing.  An example is the number of clusters per site in percolation.
Besides, there exist quantities whose scaling functions are not affected by the specific system,
but depend on the boundary conditions and need metric factors. Examples including the strength of the largest cluster (order parameter)  and wrapping probabilities in percolation,  dimensionless ratios etc. 
Furthermore, it is noted that statistical ensembles may renormalize the critical exponents~\cite{Fisher1968} and change universal values of dimensionless quantities~\cite{Hu2015}, and that the interaction anisotropy can lead to variations of universal properties~\cite{Chen2004, Hu2022, Dohm2023}.

\begin{figure}[b]
    \centering
    \includegraphics[width=0.5\textwidth]{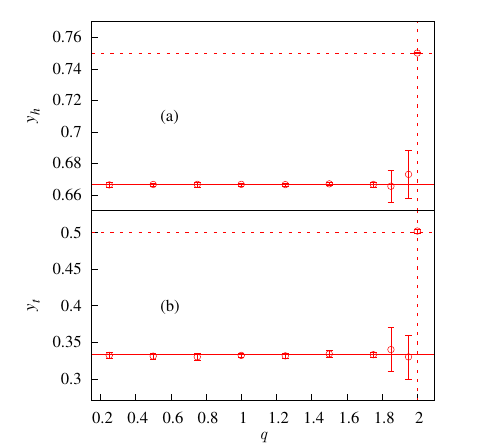}
    \caption{Numerical estimates of critical exponents for the CG Potts model with $0 < q < 2$ are consistent with theoretical predictions of $y_h=2/3$ and $y_t=1/3$. 
    With $q = 2$, the estimates are consistent with $y_h=3/4$ and $y_t=1/2$.}
    \label{yhyt}
\end{figure}

\begin{figure}[b]
    \centering
    \includegraphics[width=0.5\textwidth]{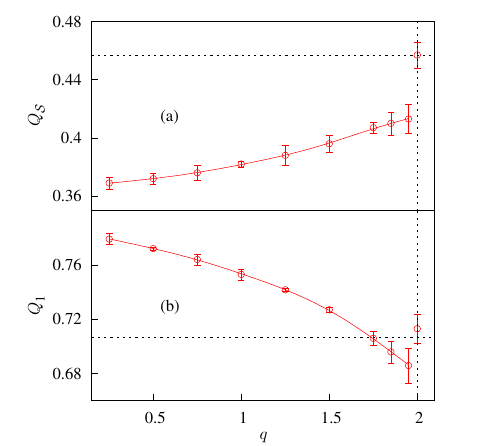}
    \caption{The dimensionless ratios $Q_{\mathcal{S}}$ and $Q_{1}$ vary continuously for $0< q <2$, 
    and show sudden changes at $q=2$.
    The smooth curves are added to guide the eye.}
    \label{Qs}
\end{figure}

The $q$-state Potts model~\cite{wupotts} is a typical model in exploring universal critical behavior.
The model is defined by the Hamiltonian $\mathcal{H}=-J\sum _{\langle ij\rangle}\delta_{\sigma_i,\sigma_j}$, 
where each site $i$ possesses a spin $\sigma_i$ taking one of $q$ states (i.e., $\sigma_i=1,2,...,$ or $q$), 
$\langle ij \rangle$ denotes a pair of neighboring sites, and $J>0$ represents the ferromagnetic coupling constant.
When $q=2$, the Potts model is equivalent to the well-known Ising model. 
The Potts spins can be mapped to random clusters~\cite{grimmett2006random} through
the Fortuin-Kasteleyn (FK) transformation~\cite{kasteleyn1969phase,fortuin1972random}.
In the transformation, each bond connecting two nearest-neighboring sites with the same spin value is independently occupied with probability $p=1-e^{-J/kT}$, 
where $k$ denotes the Boltzmann constant, $T$ represents the temperature.
The resulting graph of vertices and occupied bonds corresponds to a configuration of the Potts model in the random-cluster representation.
For a given graph $\mathcal{G}=(\mathcal{V},\mathcal{E})$, with $\mathcal{V}$ being the vertices set and $\mathcal{E}$ being the edges (bonds) set,
the Potts model in the random-cluster (RC) representation is defined by the partition function
\begin{align}
    Z=\sum_{A \subseteq \mathcal{E}} q^{k(A)}v^{|A|} \;.
\end{align}
Here $|A|$ is the number of occupied bonds, and $k(A)$ denotes the number of clusters 
(a cluster consists of vertices connected by occupied bonds; an isolated vertex is also a cluster).
The bond weight is $v=p/(1-p)$, and the cluster weight is $q$ which takes positive real numbers.
When $q \to 1$, the above Potts model reduces to standard uncorrelated bond percolation~\cite{stauffer1992introduction}.

The complete graph (CG) offers an accurate platform for finite-size approach to the mean-filed solution.
In a complete graph, there is an edge between each pair of vertices, thus each vertex interacts with all other vertices.
This makes the mean-field solutions accurate on the CG.
Critical exponents from the mean-field theory are quantitatively correct above the upper critical dimension,
and the theory offers a qualitative picture for critical phenomena in lower dimensions.
The CG Potts model has been systematically explored.
For $0 < q \leq 2$, the model shows a continuous phase transition 
and the critical point is given by $p_c={q}/{V}$~\cite{bollobas1996random}, where $V$ is the number of vertices in the CG (also called the size of the CG).
The model with $q>2$ exhibits first-order phase transitions, 
and the transition point is given by $p_c=\frac{1}{V}\frac{2(q-1)}{q-2}{\rm log}(q-1)$~\cite{wupotts,bollobas1996random}.
In the limit $q \rightarrow 0$, as the above $p_c$ corresponds to a finite $v_c/q = 1/({V-q})$,
a critical configuration of the CG Potts model in the RC representation consists of a spanning forest~\cite{jacobsen2005spanning,grimmett2006random}.

Considering universal properties of the CG Potts model with $0 < q < 2$,
from theoretical results on the thermal exponent $y_t$ and magnetic exponent $y_h$~\cite{bollobas1996random,luczak2006phase}, 
it is known that the model belongs to the universality class of standard percolation on the CG.
For finite-dimensional Potts models below the upper critical dimension,
critical exponents are usually continuous functions of $q$ and the universality of the Potts model is $q$-dependent. 
For the CG Potts model with $0 < q < 2$, the above $q$-independence of critical exponents 
can be regarded as a kind of {\it hyper-universality}.

For $q=2$, it has been observed that the CG Ising model in the RC representation strongly exhibits behaviors in the CG percolation universality class~\cite{fang2021percolation}.
For example, after excluding the largest cluster,
the Fisher exponent $\tau$ characterizing the critical cluster number density is consistent with $5/2$ in the percolation case. 
It was found that, as the size $V$ increase, the critical configuration space has a decaying percolation sector in which clusters exhibit the CG percolation scaling behaviors. 
Near criticality, besides the Ising scaling window of size $O(V^{-1/2})$,  the CG Ising model has a percolation scaling window of size O$(V^{-1/3})$.
Outside the Ising window and in the percolation window, all clusters display the same scaling behaviors as those in critical CG percolation.
Recently percolation behaviors in the CG Ising model have also been demonstrated through the loop-cluster joint model in Refs.~\cite{zhang2020loop, li2023geometric}.

From above results for the CG Potts model, it could be conjectured that the Fisher exponent $\tau = 5/2$ is hyper-universal in the range $0 < q \leq 2$.
The hyper-universality of critical exponents appear very interesting and deserves more numerical verification.
The universality of quantities besides critical exponents also deserves more investigations, 
e.g., it is not clear how dimensionless ratios behave as $q$ varies.
In this paper, we numerically demonstrate the hyper-universality of critical exponents 
and thoroughly investigate the dimensionless ratios of the CG Potts model.
We have also made some improvements to the Monte Carlo algorithms for efficient simulations.

The remainder of the paper is organized as follows.
Section~\ref{result} summarizes the main results.
Section~\ref{algorithm} describes the algorithms used in numerical simulations.
Sections~\ref{exponents} and ~\ref{ratios} present detailed results for measured quantities,
including also evidence demonstrating the efficiency of the algorithms.
Section~\ref{conclusion} contains a brief conclusion and discussion.

\section{\label{result}Main results}

Our main results include a demonstration of the hyper-universality of critical exponents, and the finding of $q$-dependence of dimensionless ratios.
They are summarized as follows, while the analyses details are presented in Sections~\ref{exponents} and ~\ref{ratios}.

Firstly, for the critical exponents, our numerical results in Fig.~\ref{yhyt} demonstrates that the magnetic renormalization exponent $y_h$ 
and the thermal renormalization exponent $y_t$ remain constant for $0 < q < 2$, i.e., they are hyper-universal. The numerical results support 
theoretical predictions~\cite{bollobas1996random,luczak2006phase} of $y_h=2/3$ and $y_t=1/3$ for $0< q < 2$, and of $y_h=3/4$ and $y_t=1/2$  for $q=2$.
Thus the model with $0 < q < 2$ are in the percolation universality class, whose critical exponents are different from the Ising universality class.
Our numerical estimates for $y_h$ come from finite-size scaling (FSS) analyses on the average of the largest-cluster size  $\mathcal{C}_1$, 
and the average of the second moment of cluster sizes $\mathcal{S}_2 = \sum_i{\mathcal{C}_i}^2$. 
To the leading order they behave as $\langle {\mathcal{C}_1} \rangle \sim V^{y_h}$
and $\chi \equiv \langle \mathcal{S}_2 \rangle/V \sim V^{2y_h-1}$.
For $y_t$, our estimates come from FSS analyses on two covariances ${\rm Cov}(\mathcal{N}_B,Q_1)$ and ${\rm Cov} (\mathcal{N}_B,Q_\mathcal{S})$~\cite{blote1995ising,deng2003simultaneous,zhang2021critical},
which both scale as $\sim V^{y_t}$ to the leading order.
Here $\mathcal{N}_B = |A|$ is the number of occupied bonds in the RC representation, 
$Q_1$ and $Q_\mathcal{S}$ are two Binder-like dimensionless ratios defined as
\begin{align}
    &Q_1=\frac{{\langle \mathcal{C}_1 \rangle}^2}{\langle {\mathcal{C}_1}^2 \rangle}, \\
    &Q_\mathcal{S}=\frac{{\langle \mathcal{S}_2 \rangle}^2}{\langle 3{\mathcal{S}_2}^2-2{\mathcal{S}_4} \rangle},
\end{align}
where $\mathcal{S}_4 = \sum_i{\mathcal{C}_i}^4$ is the fourth moment of cluster sizes.
The above two covariances are correlations between $\mathcal{N}_B$ and the two ratios, 
which are defined as~\cite{blote1995ising,deng2003simultaneous,zhang2021critical}
\begin{align}
    &{\rm Cov}(\mathcal{N}_B,Q_\mathcal{S}) = \frac{2\langle \mathcal{S}_2\mathcal{N}_B \rangle}{\langle \mathcal{S}_2 \rangle} 
    - \frac{\langle (3{\mathcal{S}_2}^2-2{\mathcal{S}_4})\mathcal{N}_B \rangle}{\langle 3{\mathcal{S}_2}^2-2{\mathcal{S}_4} \rangle} 
    - \langle \mathcal{N}_B \rangle, \nonumber \\
    &{\rm Cov}(\mathcal{N}_B,Q_1) = \frac{2\langle \mathcal{C}_1\mathcal{N}_B \rangle}{\langle \mathcal{C}_1 \rangle}              
    - \frac{\langle \mathcal{C}_1^2\mathcal{N}_B \rangle}{\langle \mathcal{C}_1^2 \rangle} - \langle \mathcal{N}_B \rangle .  
\end{align}

We have also analyzed the Fisher exponent $\tau$ and the two-arm exponent $x_2$, as defined in Sections~\ref{tau} and ~\ref{x2}, respectively.
The exponent $\tau$ is associated with the cluster number density and our numerical results show that $\tau$ is invariant for $0 < q \le 2$, 
supporting the conjecture of the hyper-universality of $\tau$ in the Introduction.
Our analyses suggest that the two-arm exponent $x_2$ is related to the number of vertices visited in the simultaneous breadth-first searches 
of the Sweeny Monte Carlo algorithm, i.e., the latter number scale as $V^{y_h-x_2} \sim \log{V}$ for $0 < q < 2$, where $y_h$ takes the percolation value $2/3$.
This indicates $x_2=2/3$ for $0 < q < 2$, demonstrating the hyper-universality of $x_2$. The above scaling also explains the high efficiency of the Sweeny algorithm.

Secondly, in contrast to the above $q$-independent critical exponents, as shown in Fig.~\ref{Qs}, we find that dimensionless ratios $Q_\mathcal{S}$ and $Q_1$ 
continuously change for $0 < q < 2$ and exhibit jumps at $q=2$. These tell that, as in other models, the dimensionless ratios are less universal 
than critical exponents in the CG Potts model. 
The sudden changes at $q=2$ is related to the fact that the CG Potts model with $0<q<2$ 
and the model with $q=2$ belong to different universality classes (as categorized by values of critical exponents $y_t$ and $y_h$).
The $q$-dependence of dimensionless ratios reflects the variation of universal scaling functions 
of cluster-size distributions, which is carefully observed in Sec.~\ref{ratios}.

\section{\label{algorithm}Algorithms}

\begin{algorithm}[b]
\caption{Swendsen-Wang-Chayes-Machta algorithm for the  CG Potts model}
\label{alg:SWCM}
Independently for each cluster $i$ ($= 1,...,k(A)$), generate a random number $P_i \in [0,1)$ \\
\uIf {$P_i<=1/q$}
{The cluster is classified as ``active"}
\Else
{The cluster is classified as ``inactive"}
Update all ``active" clusters as the bond percolation.\\
Update the information of all clusters.
\end{algorithm}

\begin{algorithm}[b]
\caption{Modified Sweeny algorithm for the CG Potts model}
\label{alg:sweeny1}
Choose an operation randomly --- adding a bond or deleting a bond\\
\uIf{Adding a bond}
{
Choose an empty bond $xy$ uniformly at random\\
\uIf {$xy$ connected via a path not including $xy$} 
{Occupy $xy$ with probability $\min \{1,P^{(+)}_{BB}\}$}
\Else
{Occupy $xy$ with probability $\min\{1,P^{(+)}_{BR}\}$}
}
\ElseIf{Deleting a bond}
{
Choose an occupied bond $xy$ uniformly at random\\
\uIf {$xy$ connected via a path not including $xy$} 
{Delete $xy$ with probability $\min \{1,P^{(-)}_{BB}\}$}
\Else
{Delete $xy$ with probability $\min\{1,P^{(-)}_{BR}\}$}
}
\end{algorithm}

Markov Chain Monte Carlo (MCMC) algorithms have proven to be effective tools for simulating the Potts model.
Two widely used MCMC algorithms are
the Swendsen-Wang-Chayes-Machta (SWCM) algorithm~\cite{swendsen1987nonuniversal,chayes1998graphical}
and the Sweeny~\cite{sweeny1983monte} algorithm.
The SWCM algorithm utilizes non-local cluster updates and exhibits high efficiency for Potts model with $q>1$.
For the Potts model with $0 <q < 1$, the Sweeny algorithm is the only known MCMC algorithm and it shows rich dynamic behaviors~\cite{peng2023sweeny}.
For bond percolation ($q=1$), a configuration can be easily generated by independently occupying each edge with the same probability.

The SWCM algorithm for CG Potts model is presented in Alg.~\ref{alg:SWCM}. 
It mainly involves two steps: 
for a given bond configuration independently deem each cluster to be ``active" with probability $1/q$ and ``inactive" otherwise;
then update all ``active" clusters as uncorrelated bond percolation with occupation probability $p=v/(v+1)$.
Hereafter this work focuses on the critical point of the CG Potts model with $0<q\le2$, thus bond percolation in the latter step has $p=p_c=q/V$.
It is noted that this $p_c$  is different from bond percolation threshold $1/V'$, where $V'$ is the number of vertices in all ``active" clusters.

For percolation in the SWCM algorithm, the bond occupation probability $p=q/V$ is rather small for large $V$.
This implies that many operations would be needed if all candidate bonds were visited to decide whether or not they are occupied.
One can use a more efficient procedure as in Refs.~\cite{blote2002cluster,deng2005monte,huang2018critical}, which is described as follows.
Considering $i$ candidates bonds, we define $P(i)\equiv(1-p)^{i-1}p$ to be the probability that the first $(i-1)$ bonds are unoccupied while the $i$th bond is occupied. 
The cumulative probability is calculated as $F(i) = \sum_{j=1}^i P(j) = 1-(1-p)^i$,
which gives the probability that the first occupied bond has its index less than or equal to $i$.
After occupying one bond $i_0$, one gets the next occupied bond $(i_0+i)$
by drawing a uniformly distributed random number $0 \leq r < 1$. 
The value of $i$ satisfies
\begin{equation}
  F(i-1) \leq r < F(i).
  \label{Eq:i}
\end{equation}
From the above equation one derives
\begin{equation}
  i = 1 + \lfloor \log(r)/\log(1-p_c) \rfloor \; .
  \label{Eq:sovle}
\end{equation}
The above process is iterated until the status of all candidate bonds are determined.
The procedure is efficient as one jumps between occupied bonds, without explicitly considering 
the occupation of all candidate bonds one by one.
This approach significantly enhances the simulation efficiency for models with small $p$. 
It is particularly advantageous for the CG model which has a large total number of bonds.

For the Sweeny algorithm, a
basic step is proposing to change the status of a randomly selected bond, 
then accepting the proposal according to the Metropolis-Hastings criterion~\cite{metropolis1953equation,hastings1970monte}.
More precisely, if $xy$ is an empty bond, 
it will be occupied with probability $\min \{1,v\}$ (resp. $\min\{1,v/q\}$) in case that two endpoints $x$ and $y$ are (resp. are not) connected via a path not including $xy$. 
If $xy$ is an occupied bond, it will be erased with probability $\min\{1,1/v\}$ (resp. $\min\{1,q/v\}$), 
in case that two endpoints of $xy$ are (resp. are not) connected via a path not including $xy$.
For this original Sweeny algorithm, a problem arises when $V$ becomes large: 
at $v=v_c={q}/({V-q})$, probabilities of updates occupying an empty bond are proportional to ${q}/({V-q})$ or ${1}/({V-q})$, which become very small for large $V$. 
This would cause a severely slowing-down of the simulations~\cite{peng2023sweeny}.
Similar to our previous work~\cite{peng2023sweeny}, we propose a modified Sweeny algorithm for CG Potts model, as depicted in Alg.~\ref{alg:sweeny1}. 

In the original Sweeny algorithm, the probability of proposing the occupation of an empty bond is high, but the corresponding acceptance probabilities are low. 
In the modified algorithm, we increase the acceptance probabilities by equalizing proposal probabilities of adding and deleting bonds.
The acceptance probabilities in Alg.~\ref{alg:sweeny1} are derived as follows.
Using $A$ and $B$ to represent two configurations that can transform into each other in an update, the detailed balance condition has the form
\begin{align}
&\pi(A){P_{\rm pro}}(A\rightarrow B){P_{\rm acc}}(A\rightarrow B)\\ \nonumber
=&\pi(B){P_{\rm pro}}(B\rightarrow A){P_{\rm acc}}(B\rightarrow A),
\end{align}
where $\pi$ is the configuration weight, $P_{\rm pro}$ and $P_{\rm acc}$ are the proposal and acceptance probabilities, respectively.
The detailed balance condition associated with the addition or deletion of a bond can be written as

\begin{align}
    &\frac{1}{\mathcal{N} - \mathcal{N}_{\rm B}(A)} \times q^{k(A)}v^{\mathcal{N}_{\rm B}(A)} \times P_{\rm acc}({A \to B}) \\ \nonumber
    =&\frac{1}{\mathcal{N}_{\rm B}(B)} \times  q^{k(B)}v^{\mathcal{N}_{\rm B}(B)} \times P_{\rm acc}({B \to A}) .  
\end{align}

Considering the case where $A\to B$ represents the occupation of a bond $xy$ ($B\to A$ represents the deletion of the bond):
when $x$ and $y$ are (resp. not) connected through a path that does not include the bond $xy$, 
one observes that $k(A) = k(B)$ (resp. $k(A) -1 = k(B)$ ) and $\mathcal{N}_{\rm B}(A)+1=\mathcal{N}_{\rm B}(B)$. 
Following the Metropolis-Hastings criterion, at $v=v_c$ the acceptance probabilities in Alg.~\ref{alg:sweeny1} are calculated as
\begin{align}
     &P^{(+)}_{BB} = \min \{ 1,\frac{\mathcal{N} - \mathcal{N}_{\rm B}}{\mathcal{N}_{\rm B}+1} \times \frac{q}{V-q} \}, \\
     &P^{(+)}_{BR} =\min \{ 1,\frac{\mathcal{N} - \mathcal{N}_{\rm B}}{\mathcal{N}_{\rm B}+1} \times \frac{1}{V-q} \}, \\ 
     &P^{(-)}_{BB} =\min \{ 1,\frac{\mathcal{N}_{\rm B}}{\mathcal{N} - \mathcal{N}_{\rm B}+1} \times \frac{V-q}{q} \}, \\
     &P^{(-)}_{BR} =\min \{ 1,\frac{\mathcal{N}_{\rm B}}{\mathcal{N} - \mathcal{N}_{\rm B}+1} \times \frac{V-q}{1} \}.
\end{align}
Given $\mathcal{N}=\frac{V(V-1)}{2}$ and $\langle \mathcal{N}_{\rm B} \rangle \simeq \frac{V}{2}$~\cite{bollobas1996random}, for large $V$ we estimate that 
\begin{align}
     &P^{(+)}_{BB} \simeq \min \{ 1, q \},\\
     &P^{(+)}_{BR} \simeq 1, \\
     &P^{(-)}_{BB} \simeq \min \{ 1,\frac{1}{q} \}, \\
     &P^{(-)}_{BR} \simeq 1.
\end{align}
Thus, the acceptance probabilities in our revised algorithm does do not decrease when $V$ becomes large. 

We note that a primary challenge of the Sweeney algorithm lies in verifying the connectivity between two endpoints of a chosen bond. 
For two-dimensional lattices, there are three common algorithms with distinct different asymptotic running-time properties: breadth-first search, union-find, and dynamic-connectivity algorithms~\cite{elcci2013efficient,elci2015algorithmic}.
For the CG Potts model, we use the method of simultaneous breadth-first searches starting at both end points of a bond $xy$, 
which is efficient as analyzed in Sec.~\ref{x2}.

\section{\label{exponents}Hyper-universality of critical exponents}
In this section, we demonstrate the hyper-universality of critical exponents for the CG Potts model. 
It is numerically verified that the renormalization exponents $y_t$, $y_h$, the exponent $\tau$
describing the cluster number density, and the two-arm exponent $x_2$ all keep invariant in the range $0<q<2$,
and that the invariance of $\tau$ extends to $q=2$.
When $0<q<2$, we also show that $x_2=y_h$ explains the efficiency of using simultaneous breath-first searches 
in the Sweeny algorithm.

\subsection{\label{ytyh}Renormalization exponents $y_t$ and $y_h$}
The exponents $y_t$ and $y_h$ have been theoretically derived for the CG Potts model with $0<q\le2$.
The model was first systematically studied in Ref.~\cite{bollobas1996random}, 
and then extended to a broad family of pseudo-critical points in Ref.~\cite{luczak2006phase}.
Defining $\varepsilon \equiv |p - p_c|/p_c$ with $p_c=q/V$, these works derived the following:
\begin{enumerate}[label=(\roman*)]
    \item For $0< q < 2$, when $p = \frac{q}{V}(1 \pm \varepsilon)$ with $\varepsilon \le O(V^{-1/3})$, the largest cluster scales as $\langle {\mathcal{C}_1} \rangle \sim V^{2/3}$;
    \item For $q = 2$, when $p = \frac{2}{V}(1 \pm \varepsilon)$ with $\varepsilon \le O(V^{-1/2})$, the largest cluster behaves as $\langle {\mathcal{C}_1} \rangle \sim V^{3/4}$.
\end{enumerate}
From finite-size scaling theory, usually the size of a critical scaling window scales as $ O(V^{-y_t})$,
and the largest cluster behaves as $\langle {\mathcal{C}_1} \rangle \sim V^{y_h}$.
Thus, when $0<q<2$, from (i), the CG Potts model belongs to the percolation universality class with $y_t=1/3$ and $y_h=2/3$;
when $q=2$, from (ii), the model falls within the Ising universality class with $y_t=1/2$ and $y_h=3/4$. 
To demonstrate the theoretical results, we numerically determined $y_h$ and $y_t$ as plotted in Fig.~\ref{yhyt}.
Simulations were performed for different sizes $V$ at the critical point.
Then the data were fitted to the formula $aV^{y_0}(1+b_1V^{y_1})+c_0$ using the least-square method, 
where $y_0>0$ is the leading scaling exponent and $y_1<0$ is the leading correction exponent.
Estimates of $y_h$ come from fitting results of $\langle \mathcal{C}_1 \rangle$ and $\langle \mathcal{S}_2 \rangle/V$,
for which $y_0=y_h$ and $2y_h-1$, respectively.
Estimates of $y_t$ come from fitting the data of covariances ${\rm Cov}(\mathcal{N}_B,Q_1)$ and ${\rm Cov}(\mathcal{N}_B,Q_\mathcal{S})$,
for which $y_0=y_t$.
Figure~\ref{yhyt} shows that our numerical results of $y_t$ and $y_h$ are consistent with theoretical predictions in last paragraph,
thus supporting the hyper-universality of $y_t$ and $y_h$ for the CG Potts model with $0<q<2$.

\begin{figure}[t]
    \centering
    \includegraphics[width=0.5\textwidth]{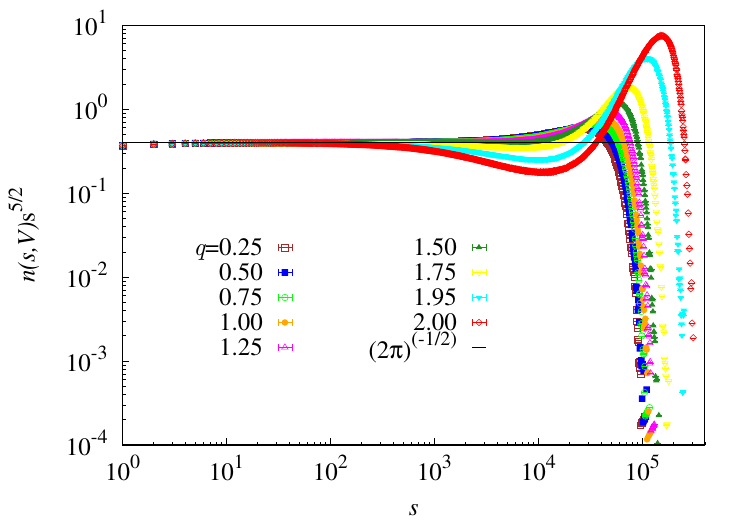}
    \caption{
	    The scaled cluster number density $n(s,V)s^{5/2}$ versus the cluster size $s$
	    for the CG Potts model with different $q$. The data were obtained from simulations
	    at fixed $V=2^{22}$.
	    The horitonzal straight line represents a theoretical prediction
	    of $n(s,V)s^{5/2}=(2\pi)^{-1/2}$ for small $s$~\cite{ben2005kinetic}.}
    \label{nst}
\end{figure}

\begin{figure}[t]
    \centering
    \includegraphics[width=0.5\textwidth]{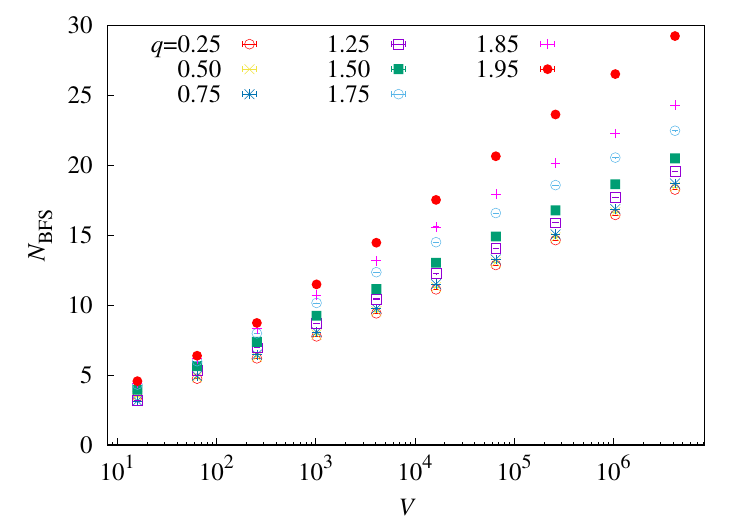}
    \caption{Numerical results of the number of vertices $N_{\rm BFS}$ that are needed to be searched 
    in the simultaneous breadth-first searches of the Sweeny algorithm. 
    The approximate linearity in the semi-log plot suggests that $N_{\rm BFS} \sim \log{V}$ for $0<q<2$.
    }
    \label{BFS}
\end{figure}  

\subsection{\label{tau}Critical exponent $\tau$}
The exponent $\tau$ describes the cluster number density $n(s,V)$.
The quantity $n(s,V)\Delta s$ gives the number of clusters with sizes in $[s, s + \Delta s]$, 
normalized by the system size $V$.
For standard percolation ($q=1$) on the CG, Ref.~\cite{ben2005kinetic} predicts that $n(s,V)$ follows a standard scaling form
\begin{align}
    n(s,V)=n_0s^{-\tau}\tilde{n}(s/V^{d_{\rm{f}}}) \qquad [\tilde{n}(x \to 0)=1],
    \label{eqns}
\end{align}
with $n_0=(2\pi)^{-1/2}$, $d_{\rm{f}}=y_h=2/3$, and $\tau = 5/2$.
For the CG Potts model with $q=2$, in Ref.~\cite{fang2021percolation} it was observed that
$\tau$ and $n_0$ have the same value as the $q=1$ model, though exponents $y_t$ and $y_h$ 
are different for the $q=2$ and $q=1$ models.
For these results, we conjecture that for other models with $q \in (0,2]$,
values of $n_0$ and $\tau$ also have the same values as the $q=1$ model. 
We plotted our numerical results of $n(s,V)$ in Fig.~\ref{nst}. 
The figure shows that, for small values of $s$, $n(s,V)s^{5/2}$ is consistent with $(2\pi)^{-1/2}$, 
which demonstrates the above conjecture. 
The deviations at large $s$ are supposed to be accounted by finite-size effects.
Thus, the critical exponent $\tau$ is hyper-universal for $0 < q \le 2$.

Results on the CG Ising model ($q=2$) can be understood by the loop-cluster (LC) joint model~\cite{zhang2020loop}.
For the CG Ising model in the loop representation, the bond density decays as $V^{-1/2}$, which means that the loop configurations 
are almost vacant in the thermodynamic limit~\cite{li2023geometric}.
Under the LC joint model, a LC algorithm passes back and forth between the loop and FK cluster representations.
The process from the loop to the FK configuration is like a conventional percolation process~\cite{li2023geometric}.
During this process, almost all loops are connected together to form the largest FK cluster,
and other FK clusters are basically standard percolation clusters. 
In other words, for the CG Ising model, only the largest cluster demonstrates the Ising property, 
while other clusters manifest percolation properties.
These observations underscore the profound intricacy and richness of universality.

\subsection{\label{x2}Two-arm exponent $x_2$}
On a finite-dimensional lattice, the $k$-arm exponent $x_k$ describes the probability that $k$ neighboring vertices
within $O(1)$ distances connect to the surface of a (hyper-) sphere of radius $R$ through $k$ distinct clusters, 
i.e., for large $R$ the probability asymptotically scales as $p_k(R) \sim R^{-x_k}$. 
For a finite-size lattice of linear size $L$, at criticality one has $p_k(L) \sim L^{-x_k}$.
When employing the Sweeny algorithm, we use the method of simultaneous breadth-first searches~\cite{deng2010some,elcci2013efficient},
which start at both endpoints of a bond $xy$ to check if $x$ and $y$ are connected via a path not including $xy$.
This method of simultaneous searches has very simple code and is effective enough for small and medium lattice sizes, 
as demonstrated in Ref.~\cite{deng2010some}. 
When $x$ and $y$ are not connected via a path not including $xy$, the two endpoints are associated with two distinct clusters 
if assuming the bond $xy$ is unoccupied, and Ref.~\cite{deng2010some} shows that the average size of the smaller cluster 
scales as $\sim L^{d_{\rm F} - x_2}$ on the square lattice , where $d_{\rm F}$ is the fractal dimension and $x_2$ is the two-arm exponent~\cite{saleur1987exact}.

\begin{figure}[b]
    \centering
    \includegraphics[width=0.5\textwidth]{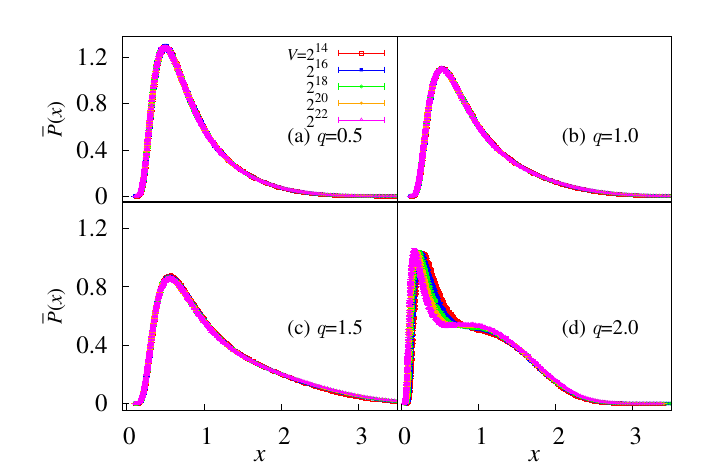}
    \caption{
    The probability distribution function $\Bar{P}(x)$ versus $x$ at (a) $q=0.5$, (b) $q=1$, (c) $q=1.5$ and $q=2$.
    The variable $x \equiv \mathcal{C}_1 / \langle \mathcal{C}_1 \rangle$, where $\langle \mathcal{C}_1 \rangle \sim V^{y_h}$
    with $y_h = 2/3$ for $0<q<2$ and $y_h=3/4$ for $q=2$.
    }
    \label{c1}
\end{figure}

\begin{figure}[b]
    \centering
    \includegraphics[width=0.5\textwidth]{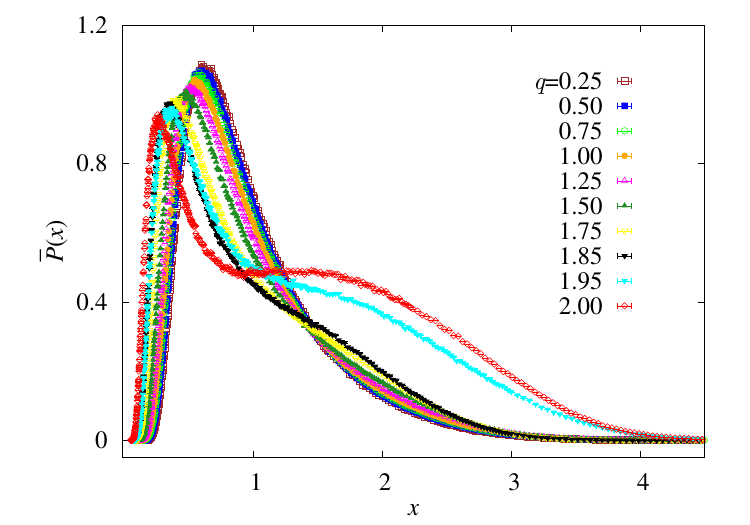}
    \caption{
    The probability distribution function $\Bar{P}(x)$ versus $x$ at $V=2^{22}$.
    The variable $x \equiv \mathcal{C}_1 / \langle \mathcal{C}_1 \rangle$, where $\langle \mathcal{C}_1 \rangle \sim V^{y_h}$
    with $y_h = 2/3$ for $0<q<2$ and $y_h=3/4$ for $q=2$.
    The function $\Bar{P}(x)$ evolves as $q$ changes. 
    }
    \label{c1nsc1}
\end{figure}

For the CG of size $V$, though the Euclidean distance cannot be defined, one can similarly define $p_k(V)$ 
as the probability that there exist $k$ pairs of vertices, where two vertices in a pair belong to a same cluster
and different pairs are in distinct clusters.
At criticality, similarly we define that the $k$-arm exponent $x_k$ describes the probability as $p_k(V) \sim V^{-x_k}$.
For $k=1$, considering that sizes of critical clusters scale as $V^{d_{\rm f}}$ (which is true for $0<q<2$), with $d_{\rm f}=y_h$ being the volume fractal dimension, 
the probability of two vertices being in the same cluster is given by $p_1(V) \sim V^{y_h-1}$, i.e., $x_1=1-y_h$.
In the limit of $V \to \infty$ on the CG, it is expected that selections of $k$ pairs of vertices become mutually independent,
which indicates $x_k=k(1-y_h)$. 

Similar to that on finite-dimensional lattices, on the CG we assume that the number of vertices $N_{\rm BFS}$ 
needed to be searched in the simultaneous breadth-first searches of the Sweeny algorithm scales as $\sim V^{y_h - x_2}$.
We measured $N_{\rm BFS}$ to determine $x_2$. From Fig.~\ref{BFS}, it is seen that $N_{\rm BFS} \sim \log V$ for $0<q < 2$
(for $q=1$, the Sweeny algorithm is not needed and we did not measure $N_{\rm BFS}$). 
The logarithmic behavior suggest that the Sweeny algorithm with simultaneous breadth-first searches 
is a very efficient method for simulating the CG Potts model.
The logarithmic behavior also indicates $x_2=y_h$ for $0 <q < 2$. 
Thus, for $0<q<2$, one has $x_2=y_h=2/3$, i.e., it is hyper-universal. 

\section{\label{ratios}$q$-dependent dimensionless ratios and the cluster-size distributions}
While critical exponents are found to be hyper-universal, 
we find that critical dimensionless ratios still depend on the parameter $q$,
as demonstrated for the ratios $Q_{\mathcal{S}}$ and $Q_1$ in Fig.~\ref{Qs}.
In this section, we present details for estimating the dimensionless ratios.
We also attribute the variation of the ratios to the change of the cluster-size distributions.

\begin{table*}
\caption{\label{table2}
Fit results for $Q_1$ at $q=2$. The simulation data were fitted to the formula $Q_1(V)=Q_{1,0}+b_1V^{y_1}+b_2V^{y_2}$ using the least-square method.
Only data with $V \ge V_{\rm min}$ were included in the fits to eliminate effects from higher-order corrections. 
``DF" is short for degrees of freedom in the fits.}

\begin{tabular}{llllllll}

\hline 
&$V_{\rm min}$  	&$Q_{1,0}$	&$y_1$ 	&$y_2$   &$b_1$ 	&$b_2$ 	&$\chi^2/{\rm DF}$ 	\\
\hline 
&$2^8$      &0.708(4)  &-0.124(9) &2$y_1$   &-0.06(3) 	&0.36(2)	&3.2/4\\ 

&$2^{10}$   &0.713(10)  &-0.111(19) &2$y_1$   &-0.10(7) 	&0.37(3)	&2.76/3\\

&$2^{8}$   &0.707(3)  &-0.16(1) &3/2$y_1$   &-0.14(6) &0.43(4)	&3.22/4\\

&$2^{10}$   &0.712(9)  &-0.14(3) &3/2$y_1$   &-0.2(1) 	 &0.47(8)	&2.74/3\\

&$2^{10}$   &0.706(4)   &-1/6  &-1/4  &-0.10(1)  &0.43(2)  &5.63/5\\ 

&$2^{12}$   &0.707(2)  &-1/6  &-1/4  &-0.13(2)  &0.45(4)  &3.87/3\\
\hline
\end{tabular} 
\end{table*}

\begin{table*}
\caption{\label{table1}
Estimates of $Q_{\mathcal{S}}$, $Q_1$, ${\rm{Ske}}(\mathcal{C}_1)$ and ${\rm{Kur}}(\mathcal{C}_1)$. 
The last column contains theoretical predictions at $q=2$ derived in the Appendix.
}
\begin{tabular}{lllllllllllll}
\hline
&$q$  	&0.25 	&0.5 	&0.75 	&1	&1.25  &1.5   &1.75   &1.85   &1.95   &2  &2 (theo.)	\\
\hline
&$Q_{\mathcal{S}}$ &0.369(4) &0.372(4) &0.3761(5) &0.382(2)	&0.388(7)	&0.396(6) &0.407(4) &0.410(8) &0.413(10) &0.457(9) & 0.4569\\ 
\hline
&$Q_1$ &0.779(3) &0.772(2) &0.764(4) &0.753(4)	&0.7417(12)	&0.727(2) &0.706(5) &0.696(8) &0.686(13) &0.713(11) & 0.7071\\ 
\hline
&${\rm{Ske}}(\mathcal{C}_1)$ &1.365(2) &1.343(2) &1.310(2) &1.27(1)	&1.21(1)	&1.128(21) &0.984(24) &0.858(18) &0.715(30) &0.417(40) & 0.4428\\ 
\hline
&${\rm{Kur}}(\mathcal{C}_1)$ &5.21(2) &5.04(1) &4.83(2) &4.59(1)	&4.29(2)	&3.90(2) &3.39(3) &3.15(3) &2.72(4) &2.40(4) & 2.4446\\ 
\hline
\end{tabular} 
\end{table*}

To estimate critical values of dimensionless ratios, the data were fitted to the formula
$Q(V)=Q_0+b_1V^{y_1}+b_2V^{y_2}$ using the least-square method, where $Q_0$ represents the value of a ratio $Q$ 
in the thermodynamic limit $V\rightarrow\infty$, $y_2<y_1<0$ are correction exponents, 
$b_1$ and $b_2$ are non-universal amplitudes.
Several fits with different values of $y_1$ and $y_2$  were performed to provide a reliable estimate.
For example, the fitting details for $Q_1$ at $q=2$ are provided in Table~\ref{table2}.
Estimated values of $Q_{\mathcal{S}}$ and $Q_1$ for $0 <q \le 2$ are summarized in Table~\ref{table1} and plotted in Fig.~\ref{Qs}.
For $q=2$, the numerical estimates are consistent with theoretical values derived in the Appendix.

\begin{figure}[t]
    \centering
    \includegraphics[width=0.5\textwidth]{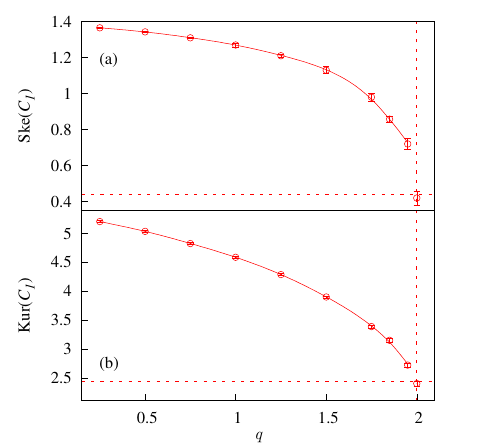}
    \caption{
    The skewness ${\rm{Ske}}(\mathcal{C}_1)$ and kurtosis  ${\rm{Kur}}(\mathcal{C}_1)$  versus $q$.
    The smooth curves are added to guide the eye, and they are not drawn to $q=2$ due to a possible discontinuity.}
    \label{skc1}
\end{figure}

\begin{figure}[t]
    \centering
    \includegraphics[width=0.48\textwidth]{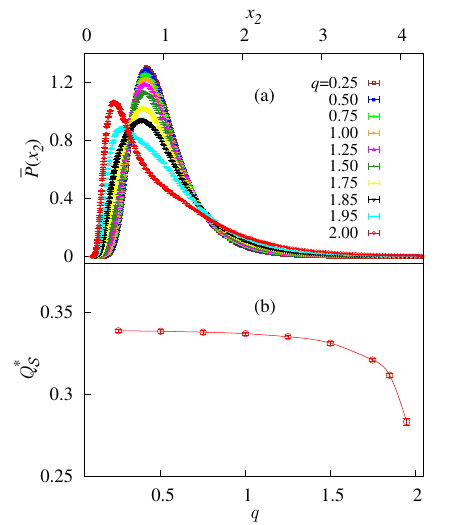}
    \caption{
    (a) The probability distribution function $\Bar{P}(x_2)$ associated with the second-largest cluster at $V=2^{22}$.
    Here the variable $x_2 \equiv \mathcal{C}_2 / \langle \mathcal{C}_2 \rangle$, where $\langle \mathcal{C}_2 \rangle \sim V^{2/3}$
    for $0<q < 2$ and $\sim V^{1/2} \log V$ for $q=2$~\cite{fang2021percolation}.
    (b) The dimensionless ratio $Q^*_{\mathcal{S}}$ versus $q$. 
    The smooth curve is added to guide the eye.
    For $q=2$, a reliable numerical estimate of $Q^*_{\mathcal{S}}$ is not available due to large finite-size corrections.}
    \label{rmc1}
\end{figure}

From Table~\ref{table1} and Fig.~\ref{Qs}, it is seen that both $Q_{\mathcal{S}}$ and $Q_1$ change continuously for $0<q<2$,
and they display jumps at $q=2$. As presented in Sec.~\ref{ytyh}, the largest cluster is govern by $y_h=3/4$ at $q=2$, 
in contrast to $y_h=2/3$ for $0<q<2$. This difference should be related to the jumps of the ratios at $q=2$.
For $0<q<2$, the continuous variation of the ratios originates from changes of the cluster-size distributions.
To illustrate the latter point, we first define $P(\mathcal{C}_1,V)d\mathcal{C}_1$ as the probability that
the largest cluster has its size between $\mathcal{C}_1$ and $\mathcal{C}_1+d\mathcal{C}_1$, 
then calculate the probability distribution function $\Bar{P}(x)dx$ with $x \equiv \mathcal{C}_1/\langle {\mathcal{C}_1} \rangle$.
The function $\Bar{P}(x)$ is depicted in Fig.~\ref{c1} for several $q$ values.
At fixed $q$, the functions asymptotically overlap at large sizes $V$, reflecting a kind of universality.
However, when comparing the functions at different $q$ values, as plotted in Fig.~\ref{c1nsc1},
one sees that $\Bar{P}(x)$ is not universal for different $q$. This explains the $q$-dependent behavior of $Q_1$, 
which characterizes the relative fluctuation of $\mathcal{C}_1$.

Besides the fluctuation, we also measured the skewness ${\rm{Ske}}(\mathcal{C}_1)$ and kurtosis  ${\rm{Kur}}(\mathcal{C}_1)$ of the distribution of $\mathcal{C}_1$, defined as
        \begin{align}
            &{\rm{Ske}}(\mathcal{C}_1)={\mu_3}/{\mu_2^{3/2}},\\
            &{\rm{Kur}}(\mathcal{C}_1)={\mu_4}/{\mu_2^2},
        \end{align}
where $\mu_i(\mathcal{C}_1) = \langle {(\mathcal{C}_1-\langle \mathcal{C}_1 \rangle)}^i\rangle$. 
The skewness quantifies the asymmetry of the probability distribution of a random variable relative to its mean, 
while the kurtosis measures the ``tailedness" of the distribution.
Results for ${\rm{Ske}}(\mathcal{C}_1)$ and ${\rm{Kur}}(\mathcal{C}_1)$ are summarized in Table~\ref{table1} 
and plotted in Fig.~\ref{skc1}. It is seen that both quantities decrease with increasing $q$.
At $q=2$, from results on $Q_1$, it could be expected that 
${\rm{Ske}}(\mathcal{C}_1)$ and ${\rm{Kur}}(\mathcal{C}_1)$ also change discontinuously.
Comparing Fig.~\ref{Qs}(b) and Fig.~\ref{skc1}, the discontinuity would be much smaller for the latter two quantities if it were there.

Finally, we ask that, after excluding the largest cluster, whether one could still observe significant change of the size distributions as $q$ varies.
We observed the size of the second-largest cluster $\mathcal{C}_2$,
and a modified dimensionless ratio $Q^*_{\mathcal{S}}$ defined as
\begin{align}
     &Q^*_{\mathcal{S}}=\frac{{\langle \mathcal{S}^*_2 \rangle}^2}{\langle 3{\mathcal{S}^*_2}^2-2{\mathcal{S}^*_4} \rangle},
\end{align}
where the modified moments are $\mathcal{S}^*_k = \mathcal{S}_k - \mathcal{C}_1^k$.
As plotted in Fig.~\ref{rmc1}(a), we find that the distribution of $\mathcal{C}_2$ varies significantly as $q$ changes.
From Fig.~\ref{rmc1}(b), one sees that the ratio $Q^*_{\mathcal{S}}$ decreases with increasing $q$,
while $Q_{\mathcal{S}}$ being monotonically increasing with $q$ in Fig.~\ref{Qs}.
Thus, excluding the largest cluster does not eliminate the $q$-dependent variation of the size distributions,
though it causes their obvious changes.

\section{\label{conclusion}CONCLUSION AND DISCUSSION}
In summary, by MCMC simulations and FSS analyses, we have studied universal properties of 
the CG Potts model with $0 < q \le 2$. The critical exponents are numerically demonstrated to 
be hyper-universal, e.g., exponents $y_t$, $y_h$ and $x_2$ are invariant for $0 < q < 2$, 
$\tau$ is invariant for $0 < q \le 2$. However, $q$-dependent variations are still observed 
in Binder-like dimensionless ratios, which are attributed to changes of the distribution functions
of the cluster sizes. In simulations, we have implemented an improved Sweeny algorithm,
and verified the high efficiency of using simultaneous breadth-first searches in the algorithm.

The above results on the CG represent mean-field behaviors of the Potts model.
We suppose that the behaviors could also be observed for the Potts model with $0 < q \le 2$ above the upper critical dimension.
The percolation model has its upper critical dimension as $d_c = 6$~\cite{chayes1987upper},
and the Ising model has two upper critical dimensions $d_c=4$ and $6$~\cite{fang2022geometric}.
Simulations of the Potts model above $d_c$ are needed to check if similar behaviors indeed present.
For these simulations above $d_c$, the improved Sweeny algorithm with simultaneous breadth-first searches is also expected to be highly efficient.
To analyze results above $d_c$, unlike for the models on the CG, the influence of boundary conditions could be vital in the FSS~\cite{fang2020, lu2024, young2024}.

\begin{acknowledgments}
    This work was supported by the National Natural Science Foundation of China under Grants No.~12275263 (Y.D.) and {No.~12375026} (H.H.), 
    the Innovation Program for Quantum Science and Technology under Grant No.~2021ZD0301900, 
    and the Natural Science Foundation of Fujian Province of China under Grant No.~2023J02032.
\end{acknowledgments}

\appendix*
\section{\label{appendix-b}Theoretical calculations of Binder-like ratios at $q=2$.}
In this appendix we give the theoretical predictions of $Q_{\mathcal{S}}$, $Q_1$, ${\rm{Ske}}(\mathcal{C}_1)$ and ${\rm{Kur}}(\mathcal{C}_1)$ for $q=2$.
The last three quantities can be derived from the probability distribution $P(\mathcal{C}_1,V)$.
For this purpose we reformulate the distribution function by using $\tilde{P}(\tilde{x})d\tilde{x}=P(\mathcal{C}_1,V)d\mathcal{C}_1$, 
where $\tilde{x} \equiv \mathcal{C}_1/V^{d_{\rm f}}$ with $d_{\rm f}=y_h=3/4$ at $q=2$.
Following Eq.~(32) in Ref.~\cite{luczak2006phase} and Eq.~(3) in Ref.~\cite{fang2021percolation}, as $V \to \infty$ the limiting distribution function is
\begin{align}
    \tilde{P}(\tilde{x}) = \frac{{\rm{exp}}(-\tilde{x}^4/12)}{\int_0^{\infty}\rm{exp}(-t^4/12)dt} \;.
\end{align}
The $k$th moment ${\langle {\mathcal{C}_1}^k \rangle}$ is 
\begin{align}
    {\langle {\mathcal{C}_1}^k \rangle} = V^{3k/4} \frac{\int_0^{\infty}\tilde{x}^k{\rm{exp}}(-\tilde{x}^4/12)d\tilde{x}}{\int_0^{\infty}\rm{exp}(-t^4/12)dt} \;.
\end{align}
Then the Binder-like ratios can be calculated as
\begin{align}
    Q_1 &= \frac{{\langle {\mathcal{C}_1} \rangle}^2}{{\langle {\mathcal{C}_1}^2 \rangle}} \simeq 0.707107 \;, \\
    {\rm{Ske}}(\mathcal{C}_1) &= \frac{{\langle {\mathcal{C}_1}^3 \rangle}-3{\langle {\mathcal{C}_1}^2 \rangle}{\langle {\mathcal{C}_1} \rangle}+2{\langle {\mathcal{C}_1} \rangle}^3}{({\langle {\mathcal{C}_1}^2 \rangle}-{\langle {\mathcal{C}_1} \rangle}^2)^{3/2}} \simeq 0.442787 \;, \\
    {\rm{Kur}}(\mathcal{C}_1) &= \frac{{\langle {\mathcal{C}_1}^4 \rangle}-4{\langle {\mathcal{C}_1}^3 \rangle}{\langle {\mathcal{C}_1}\rangle}+6{\langle {\mathcal{C}_1}^2 \rangle}{\langle {\mathcal{C}_1} \rangle}^2-3{\langle {\mathcal{C}_1} \rangle}^4}{({\langle {\mathcal{C}_1}^2 \rangle}-{\langle {\mathcal{C}_1} \rangle}^2)^2} \\ \nonumber
    &\simeq 2.44465    \;.
\end{align}

As for the $Q_{\mathcal{S}}$, for the CG Potts model with $q=2$, one has $Q_{\mathcal{S}}=Q_m$, 
where $Q_m \equiv {{\langle m^2 \rangle^2}}/{{\langle m^4 \rangle}}$ is the Binder-like ratio for the magnetization $m$.
From Eq.(A.14) in Ref.~\cite{luijten1997interaction}, one gets
\begin{align}
    Q_m &= 4[\frac{\Gamma(\frac{3}{4})}{\Gamma(\frac{1}{4})}]^2+\frac{16}{5}\sqrt{3}[\frac{\Gamma(\frac{3}{4})}{\Gamma(\frac{1}{4})}]^3\frac{1}{\sqrt{V}}+\mathcal{O}(\frac{1}{V}) \\ \nonumber
      &\simeq 0.456947 + 0.214002\frac{1}{\sqrt{V}} + \mathcal{O}(\frac{1}{V}).
\end{align}
Thus we have $Q_{\mathcal{S}} = Q_m \simeq 0.456947$ in the limit $V \to \infty$.

\bibliographystyle{apsrev4-2}
\bibliography{main.bib}
\end{document}